# From Quantum Deformations of Relativistic Symmetries to Modified Kinematics and Dynamics[*]

Jerzy Lukierski

Institute for Theoretical Physics, University of Wrocław,
50-205 Wrocław, Poland, *e-mail:lukier@ift.uni.wroc.pl*

October 11, 2018

## Abstract

We present a short review describing the use of noncommutative space-time in quantum-deformed dynamical theories: classical and quantum mechanics as well as classical and quantum field theory. We expose the role of Hopf algebras and their realizations (noncommutative modules) as important mathematical tool describing quantum-deformed symmetries: quantum Lie groups and quantum Lie algebras. We consider in some detail the most studied examples of noncommutative space-time geometry: the canonical and $\kappa$-deformed cases. Finally we briefly describe the modifications of Einstein gravity obtained by introduction of noncommutative space-time coordinates.

# Contents

# 1 Introduction

The idea that it should be useful to introduce noncommutative space-time coordinates comes already from the thirties of past century (see e.g. [1]); the first published papers on the noncommutativity of space-time appeared in forties (see e.g. [2]). Subsequently in eighties there was formulated the mathematics of noncommutative geometries (see e.g. [3, 4]) as well were given foundations of the theory of quantum groups (see e.g. [5]–[7]). The papers in which firstly appeared the quantum deformations of Poincaré symmetries (Poincaré algebras and Poincaré group) were published about 20 years ago [8]–[10].

From the physical point of view the issue of noncommutative structure of space-time is linked with short distance behaviour of phenomena or equivalently with the domain of extremely high energies. The notion of classical (pseudo) Riemannian manifold as describing the totality of physical phenomena, including quantum - mechanical and gravitational effects, appears to be not valid. If one combines together the dynamics of general relativity (Einstein equations) with the uncertainty principle of quantum mechanics, one can demonstrate that the exact localization of classical space-time events is not possible. Measuring the position of coordinate $x$ with accuracy $\Delta x$ means adding during the measurement process to small volume $(\Delta x)^3$ the big energy $E \sim \frac{1}{\Delta x}$. Due to Einstein equations such measurement - induced modification of local energy density creates the Planckian black holes with Schwarzschild radius $R \simeq \lambda_p \frac{E}{M_p}$, where $\lambda_p$ is the Planck length ($\sim 10^{-33}$cm) and $M_p$ describes the Planck mass ($\sim 10^{19}$GeV in energy units.) By formalizing above considerations Doplicher, Fredenhagen and Roberts (DFR) [11] introduced new algebraic uncertainty relations between the relativistic space-time coordinates, which are caused by the quantum gravity effects

$$[\widehat{x}_\mu, \widehat{x}_\nu] = \frac{i}{M_p^2}\theta^{(0)}_{\mu\nu}, \qquad \theta^{(0)}_{\mu\nu} = -\theta^{(0)}_{\nu\mu} \qquad [\theta^{(0)}_{\mu\nu}, \widehat{x}_\rho] = 0 \,. \qquad (1.1)$$

The deformation (1.1) is called usually the canonical one, under assumption that $\theta^{(0)}_{\mu\nu}$ is a numerical matrix. It should be added that little later, by considering the quantization of ten-dimensional free string, Seiberg and Witten [12] did demonstrate that the $D$-brane coordinates located at the end points of the string satisfy as well the relation (1.1), however in general case with point - dependent noncommutativity $(\theta^{(0)}_{\mu\nu} \to \theta_{\mu\nu}(\widehat{x}))$. One can postulate therefore more general relations

$$[\widehat{x}_\mu, \widehat{x}_\nu] = \frac{i}{M_p^2}\theta_{\mu\nu}(M_p x) = \frac{i}{M_p^2}\theta^{(0)}_{\mu\nu} + \frac{i}{M_p}\theta^{(1)\rho}_{\mu\nu}\widehat{x}_\rho \\ + i\theta^{(i)\rho\tau}_{\mu\nu}\widehat{x}_\rho\widehat{x}_\tau + \ldots \qquad (1.2)$$

The first term in (1.2) describes the canonical (DFR) deformation, the second one introduces so-called Lie-algebraic deformation and the last one the quadratic deformation. It should be added that due to proper insertions of mass parameter $M_p$ the tensors $\theta^{(0)}_{\mu\nu}, \theta^{(1)\rho}_{\mu\nu}, \theta^{(2)\rho\tau}_{\mu\nu}$ are sets of dimensionless numbers. Among Lie-algebraic deformations the earliest one is the so-called $\kappa$-deformation [10], introducing in its standard version noncommutative time coordinate by means of the following relations

$$[\widehat{x}_0, \widehat{x}_i] = \frac{i}{\kappa}\widehat{x}_i, \qquad [\widehat{x}_i, \widehat{x}_j] = 0. \qquad (1.3)$$

From second relation (1.3) follows that the nonrelativistic physics needs not to be changed in $\kappa$-deformed theory;[1] only for very high energies the first relation (1.3) implies the modification of Lorentz boost transformations and new relativistic kinematics.

It should be added still another reason for studying noncommutative spacetimes: the quantum gravity; when it is formulated as four dimensional quantum field theory with canonically quantized Einstein action, is not renormalizable in standard perturbative framework. There were tried various ways in order to overcome this difficulty: supplementing additional space dimensions (Kaluza-Klein idea), introducing additional fields and new symmetries (e.g. supersymmetries), or proposing as fundamental the extended objects (strings or in general case $p$-branes). Replacing the commutative coordinates by its noncommutative counterparts is also one of possible theoretical concepts, which it was hoped could lead to new quantum gravity formalism without devastating nonrenormalizable infinities. Such an approach has been proposed as well in the framework of canonical quantization scheme of Einstein gravity by introducing new nonlocal parametrization based on invariant loop variables which became the fundamental geometric objects. Unfortunately, till present time the nonlocal structure of quantum loop gravity still did not produce an effective way for dealing with calculational challenges of quantum gravity.

The formalism of noncommutative geometry historically was developed in two complementary ways: on one side there was studied the formalism of quantum groups and their representations (noncommutative modules) [5]–[7], in other way the noncommutative geometry was considered as the extension of functional analysis and differential calculus to the noncommutative algebras [3, 4]. Mathematical physics with its stress on the consideration of symmetries as basic concepts tends to be closer to the first approach where the notion of deformed symmetries is a primary one; in this minireview we shall follow such a way.

---
[1] In relation (1.3) the parameter $\kappa$ is a fundamental mass parameter, which should be phenomenologically related with the Planck mass $M_p$.

Firstly, in Sect. 2 we shall present the mathematical preliminaries, in particular the properties of Hopf algebras and their representations. In Sect. 3 we shall derive, starting from the noncommutative examples of space-times (see (1.2)), the corresponding deformed quantum space-time symmetries. We stress that for quantum Lie groups and quantum Lie algebras, which are the noncommutative extensions of classical Lie groups and classical Lie algebras, the basic algebraic tool is provided by Hopf algebras (see e.g. [13]). In this talk we shall consider in some detail only two examples of quantum deformations of relativistic (Poincaré) symmetries: the canonical one (see (1.1)) and the $\kappa$-deformation (see (1.3))[2]. It appears that the canonical deformation is much milder; in particular the canonically deformed Poincaré-Hopf algebra provides an example of so-called twisted quantum algebra. In effect, the theory of irreducible representations of canonically deformed and standard Poincaré algebras are the same; the difference due to deformation occurs only if we introduce the reducible tensor product representations. In the case of $\kappa$-deformations of Poincaré algebra the change of physical picture is more substantial: also the irreducible Poincaré representations are modified, what is linked with the deformation of relativistic kinematics (see Sect. 4). In Sect. 5 we shall consider briefly the canonical (see e.g. [16]–[20]) as well as the $\kappa$-deformed (see e.g. [21]–[24]) noncommutative field theory. We shall present as the basic building block in noncommutative QFT the notion of deformed quantized free fields. Usually it is conjectured that such noncommutative quantum fields should be introduced if we take algebraically into consideration the effects of quantum gravity. In Sect. 6 we briefly outline the recent noncommutative modifications of Einstein gravity. In last Sect. 7 we present a brief outlook.

## 2  Hopf algebras and the description of quantum symmetries

The continuous symmetry transformations of a physical system which is described by finite-dimensional vector space are represented by matrix Lie group $G$ given by the set of invertible matrices $T = (t_{ij})$, classified further by the choice of additional conditions (e.g. $T^+T = TT^+ = 1$ for unitary groups, where $t_{ij}^+ = t_{ji}^*$). Local (infinitesimal) description of undeformed continuous symmetries is given by the Lie algebra $\widehat{g} = (I_A)$ where $[I_A, I_B] = f_{AB}^{\ \ C} I_C$ ($A = 1 \ldots N = \dim G$). For matrix groups the Lie algebra generators are described by matrices $I_A = (I_A)_{ij}$

---

[2]For other deformations of Minkowski spaces, in particular the quadratic ones see [14, 15].

which provide group matrix $T$ via the exponential map

$$T(\alpha_1 \ldots \alpha_N) = \exp\left(\sum_{i=1}^{N} \alpha_i I_i\right), \tag{2.1}$$

where the set of numbers $(\alpha_1 \ldots \alpha_N)$ describe the continuous group parameters.

In order to describe the new symmetries of physics based on finite-dimensional noncommutative geometries (e.g. noncommutative space-time) one should introduce suitable extensions of Lie groups and Lie algebras, described by quantum matrix Lie groups and quantum Lie algebras. These new mathematical objects are examples of Hopf algebras, which permit to introduce matrix groups with noncommutative entries.

The Hopf algebra $H = (A, m, A, S, \varepsilon))$ is the unital associative algebra $A$ with multiplication $m: A \otimes A \to A$ ($m(a \otimes b) = ab$, $a, b \in A$) and the coalgebraic structure described by the coproduct $\Delta : A \to A \otimes A$ (in shorthand notation $\Delta(a) = a_{(1)} \otimes a_{(2)}$) satisfying the property

$$\Delta(a)\Delta(b) = \Delta(ab). \tag{2.2}$$

Further Hopf-algebraic structure contains the notion of coinverse (antipode) $S : A \to A$ and counit $\varepsilon : A \to C$.

The classical matrix Lie groups and Lie algebras have the universal formulae for the coproduct

$$G: \quad \Delta^{(0)} t_{ij} = t_{ik} \otimes t_{kj} \quad (\Delta^{(0)}(T) = T \dot{\otimes} T) \tag{2.3}$$

$$\widehat{g}: \quad \Delta^{(0)}(I_A) = I_A \otimes 1 + 1 \otimes I_A \quad (A = 1, 2 \ldots \dim \widehat{g}). \tag{2.4}$$

The co-algebraic structure due to the relation (2.2) can be extended to the algebra of polynomial functions $f(T)$ on the group $G$ and to the enveloping algebra $U(\widehat{g})$ of Lie algebra $\widehat{g}$.

The quantum extension $G \to \widehat{G}$ of matrix groups permits to introduce the noncommutative entries $\widehat{T} = (\widehat{t}_{ij})$ satisfying the quadratic equation $R\widehat{T}\widehat{T} = \widehat{T}\widehat{T}R$, or more explicitly

$$R_{ijkl}\,\widehat{t}_{km}\,\widehat{t}_{ln} = \widehat{t}_{jl}\,\widehat{t}_{lk}\,R_{klmn}, \tag{2.5}$$

where the quantum $R$-matrix $R = R^{(1)} \otimes R^{(2)} \in A \otimes A$ satisfies the quantum Yang-Baxter equation

$$R_{12}R_{13}R_{23} = R_{23}R_{13}R_{13}, \tag{2.6}$$

where $R_{12} = R^{(1)} \otimes R^{(2)} \otimes 1$, $R_{13} = R^{(1)} \otimes 1 \otimes R^{(2)}$ and $R_{23} = 1 \otimes R^{(1)} \otimes R^{(2)}$.

The relation (2.6) follows from the associativity of the algebra of quantum matrices $\widehat{T}$. It should be stressed that for quantum matrix groups the classical coproduct formula (2.3) remains unchanged:

$$\widehat{G}: \quad \Delta(\widehat{t}_{ij}) = \widehat{t}_{ij} \otimes \widehat{t}_{kj}. \tag{2.7}$$

The quantum counterpart of the pairs of classical groups $G$ and Lie algebras $\widehat{g}$ linked by the exponential map (2.1) is the pair $H, \widetilde{H}$ of dual Hopf algebras related by the Hopf-algebraic duality relations $(a, b \in H, \widetilde{a}, \widetilde{b} \in \widetilde{H})$

$$\begin{aligned}
\langle \Delta(\widetilde{a}), a \otimes b \rangle &= \langle \widetilde{a}, a \cdot b \rangle, \\
\langle \widetilde{a} \otimes \widetilde{b}, \Delta(a) \rangle &= \langle \widetilde{a}\widetilde{b}, a \rangle, \\
\langle 1, a \rangle = \varepsilon(a), &\quad \langle \widetilde{a}, 1 \rangle = \varepsilon(\widetilde{a}),
\end{aligned} \qquad (2.8)$$

where the nondegenerate scalar product $\langle \widetilde{a} | a \rangle$ describes the evaluation map $\widetilde{a} \in \widetilde{H}$ calculated on element $a \in H$.

From relations (2.8) follows that duality links the multiplication/comultiplication sectors in Hopf algebras $H$ and $\widetilde{H}$, i.e. one can draw the following figure, where $\longleftrightsquigarrow$ denotes duality relation

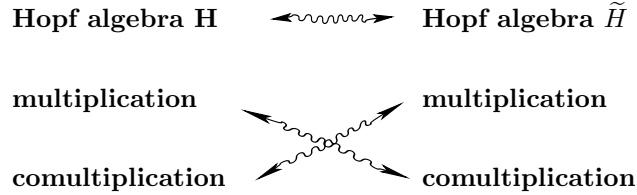

Figure 1: Hopf-algebraic duality links.

One can supplement Fig. 1 by the following comments:

(i) The variety of quantum matrix groups described by the solutions of YB equation (2.6) corresponds to the variety of choices of coproducts for the classical Lie algebras. These various coproducts are described in lowest order of deformation parameter by so-called classical $r$-matrices.

(ii) In the case of quantum Lie algebras one can choose different algebra bases. There are the ones with more complicated (more deformed) algebraic sector and simpler (less deformed) coalgebraic formulae - and vice versa: with main part of deformation present in coalgebraic relations. In particular, one can find a special basis of quantum Lie algebra in which its algebraic sector is described by classical Lie-algebra and the whole "quantum nature" is contained in coproduct formulae. The opposite statement, that whole deformation of the quantum Lie algebra is put in the algebraic sector (i.e. the coproduct can be chosen primitive, as in (2.4)) is not true - all quantum-deformed Lie algebras have non-primitive (non-cocommutative) coproducts. It should be added however that the freedom of choice of the bases of quantum Lie-algebras is only a mathematical property. In physical context, for example if the symmetries include the time translations, in principle only one choice of basis of time translation generator can be promoted to the physical Hamiltonian - it is the choice which describes shifts in physical time variable.

(iii) Summarizing in Hopf-algebraic description both algebra and coalgebra are equally important. Introducing e.g. only nonlinear modification of Lie-algebraic relations without providing coproducts does not inform us whether we deal with quantum symmetries.

The coproduct and coalgebraic structure is necessary if we wish to consider tensor product representations. The representation module $V$ of Hopf algebra $H$ is a vector space with the action $\triangleright : A \otimes V \to V$ ($a \triangleright v \in V$, where $v \in V$) satisfying the conditions (see e.g.[13])

$$ab \triangleright v = a \triangleright (b \triangleright v), \qquad (2.9)$$
$$a \triangleright vw = (a_{(1)}v) \cdot (a_{(2)}w). \qquad (2.10)$$

From (2.9) follows that the map $\triangleright$ describes a representation of $A$, and relation (2.10) implies that noncommutativity of the representation module $V$ is uniquely related with the coproduct of $H$. If we have a pair of modules $V, W$ ($v \in V, w \in W$) one obtains consistent braided tensor product $\widetilde{\otimes}$ if it satisfies the relation

$$a \triangleright (v \widetilde{\otimes} w) = a_{(1)}v \widetilde{\otimes} a_{(2)}w, \qquad (2.11)$$

where $\otimes \to \widetilde{\otimes}$ describes the deformation of symmetric standard tensor product $V \otimes W$. From (2.11) follows that because for quantum symmetry we get that $a_{(1)} \neq a_{(2)}$, the flipped standard tensor product $((a \otimes b) \circ (c \otimes d) := ac \otimes bd)$

$$(V \otimes W)^T = \tau_0 \circ (V \otimes W) = W \otimes W \qquad (2.12)$$

ceases to be a representation (this follows equivalently from the relation $[\Delta(a), \tau_0] \neq 0$). In order to introduce in explicit way the properly transposed representations we should assume that in the tensor product of Hopf algebras exists an invertible element $\widehat{R} \subset A \otimes A$ called universal $R$-matrix which satisfies the formula

$$\Delta^T(a) = \tau_0 \circ \Delta(a) \circ \tau_0^{-1} = \widehat{R}^{-1} \circ \Delta(a) \circ \widehat{R} \qquad \forall a \in A. \qquad (2.13)$$

If $\widehat{R}$ satisfies also the relations

$$(\Delta \otimes 1)\widehat{R} = \widehat{R}_{13} \widehat{R}_{23} \qquad (1 \otimes \Delta)\widehat{R} = \widehat{R}_{13} \widehat{R}_{12}, \qquad (2.14)$$
$$(1 \otimes S)\widehat{R}^{-1} = \widehat{R} \qquad (\varepsilon \otimes 1)\widehat{R} = (1 \otimes \varepsilon)\widehat{R} = 1, \qquad (2.15)$$

the Hopf-algebra $H = (A, m, \Delta, S, \varepsilon, \widehat{R})$ is called a quasitriangular Hopf-algebra. In such a case the transposed representation should be defined via $R$-matrix as follows

$$(V \widetilde{\otimes} W)^T = \widehat{R} \circ (W \widetilde{\otimes} V) = \widehat{R} \circ \tau_0 \circ (V \widetilde{\otimes} W). \qquad (2.16)$$

Because from (2.13) follows that

$$[\Delta(a), \widehat{R}\,\tau_0]_\circ = 0 \qquad \forall a \in A, \tag{2.17}$$

the definition (2.16) is consistent with the action (2.11) of $H$ on the category of tensor products. In such a way the $R$-matrix (see (2.16)) enters into the definition of braided tensor product $V\widetilde{\otimes}W$.

A subclass of quasitriangular Hopf algebras are described by the triangular Hopf algebras, for which one can define a twist factor $\widehat{F}$ ($\widehat{F} \in A \otimes A$), which satisfies the relation

$$\widehat{R} := \widehat{F}^{\,T} \circ \widehat{F}^{\,-1}. \tag{2.18}$$

The twist factor $\widehat{F}$ describes the deformation of classical Lie algebra and produces the following change of primitive coproduct $\Delta^{(0)}$ (see (2.4))

$$\Delta^F = \widehat{F} \circ \Delta^{(0)} \circ \widehat{F}^{\,-1}. \tag{2.19}$$

The antipode transforms as well as follows

$$S^F = \widehat{\chi}^{\,F} S^{(0)} \left(\widehat{\chi}^{\,(F)}\right)^{-1}, \tag{2.20}$$

where

$$\widehat{\chi}^{\,F} = F_{(1)} \cdot S\left(F_{(2)}\right) \qquad \Delta(\widehat{F}) = \widehat{F}_{(1)} \otimes \widehat{F}_{(2)}. \tag{2.21}$$

If the twist $\widehat{F}$ is a two-cochain twist, satisfying the relations

$$\widehat{F}_{12}(\Delta \otimes 1)\widehat{F} = \widehat{F}_{23}(1 \otimes \Delta)\widehat{F} \qquad (\varepsilon \otimes 1)\widehat{F} = 1 \otimes 1, \tag{2.22}$$

where $F_{23} = 1 \otimes F_{(1)} \otimes F_{(2)}$ and $(\Delta \otimes 1)F = \Delta(F_{(1)}) \otimes F_{(2)}$, the universal $\widehat{R}$-matrix described by formula (2.18) is also quasitriangular (see (2.14–2.15)).

The algebra sector of twisted classical Hopf-Lie algebras is not modified. Therefore, in the case of twisted relativistic symmetries the irreducible Poincaré algebra realizations (e.g. one-particle sector in QFT) remain unchanged; the deformation enters only into the description of tensor product representations. Using more physical terminology twisted deformation enters only into the composition formulae for the multi-particle states in terms of one-particle states. That sort of modification can be described in deformed QFT by modified algebras of field oscillators.

# 3 From noncommutative space-time to quantum relativistic symmetries

The most direct manifestation of noncommutative geometry in physics is the appearance of quantum space-time[3]. In standard mathematical approach one considers the deformed (quantum) space-time symmetry group as a primary objects and one studies noncommutative space-time as its fundamental representation module. It is however instructive to consider the opposite way - to see how various models of quantum space-time lead to the appearance of different quantum groups.

---
[3] One can consider deformation of relativistic space-time as well as of the nonrelativistic one, with Galilean time as scalar evolution parameter (see e.g. [25, 26]).

We shall consider three general classes of deformations

## 3.1 Canonical noncommutative space-time and canonical quantum Poincaré algebra

Let us consider the noncommutative space-time (1.1), with constant commutator of noncommutative coordinates $\widehat{x}_\mu$. In such a space-time the translations

$$\widehat{x}_\mu{}' = \widehat{x}_\mu + a_\mu \tag{3.1}$$

remain classical because if we assume the covariance of relations (1.1), i.e.

$$[\widehat{x}_\mu{}', \widehat{x}_\nu{}'] = \frac{i}{\kappa^2} \theta^{(0)}_{\mu\nu} \tag{3.2}$$

it follows that

$$[a_\mu, a_\nu] = [\widehat{x}_\mu, a_\nu] = 0. \tag{3.3}$$

If we assume undeformed Lorentz symmetries, the presence of constant tensor $\theta_{\mu\nu}$ in (1.1) breaks Lorentz invariance (in nondegenerate case of $\theta_{\mu\nu}$ matrix to $O(2) \times O(1,1)$ subgroup). It appears that in such approach the noncommutativity parameters describe Lorentz symmetry breaking. However in last decade it was realized that when we treat the relation (1.1) as describing the noncommutative representation module of suitably chosen quantum Poincaré group, these relations become covariant (see e.g. [17]). From the relation (see also (2.9))

$$M_{\mu\nu} \triangleright [\widehat{x}_\rho, \widehat{x}_\tau] = 0, \tag{3.4}$$

follows that for the canonically deformed Poincaré algebra one should introduce the nonprimitive coproducts of $M_{\mu\nu}$

$$\Delta(M_{\mu\nu}) = \Delta^{(0)}(M_{\mu\nu}) - \frac{1}{\kappa^2} \theta^{(0)\rho}_\mu P_\nu \wedge P_\rho. \tag{3.5}$$

Such coproducts are obtained[4] if we deform the classical Poincaré-Hopf algebra by the following twist

$$\widehat{F} = \exp\left(\frac{i}{\kappa^2} \theta^{(0)}_{\mu\nu} P^\mu \otimes P^\nu\right). \tag{3.6}$$

Using (2.19) we obtain from (3.6) the formula (3.5).

---
[4]The coproducts $\Delta(P_\mu)$ are not changed.

The presence of twist $\widehat{F}$ permits to introduce the explicit multiplication rule (star product) describing deformed product of functions on classical Minkowski space $x_\mu$ [27, 17, 28]

$$f(x) \star g(x) = m \left( \widehat{F}^{-1} \circ f(x) \otimes g(x) \right). \tag{3.7}$$

Star product (3.7) via homomorphic map (called Weyl map)

$$f(\widehat{x})g(\widehat{x}) \xrightarrow{W} f(x) \star g(x) \tag{3.8}$$

represents the noncommutative multiplication in the algebra of functions $f(\widehat{x})$. For twist (3.6) the star product formula (3.7) is provided by the Moyal product

$$f(x) \star_\theta g(x) = f(x) \exp\left\{ \frac{i}{\kappa^2} \overleftarrow{\partial}^\mu \theta^{(0)}_{\mu\nu} \overrightarrow{\partial}^\nu \right\} g(x), \tag{3.9}$$

which is obtained from (3.6)–(3.7) if we insert the differential realization of classical Poincaré algebra generators

$$P_\mu = \frac{1}{i} \partial_\mu \qquad M_{\mu\nu} = \frac{1}{i}(x_\mu \partial_\nu - x_\nu \partial_\mu). \tag{3.10}$$

In particular one can confirm by putting $f = x_\mu, g = x_\nu$ that ($[A,B]_\star := A \star B - B \star A$)

$$[x_\mu, x_\nu]_{\star_\theta} = \frac{i}{\kappa^2} \theta^{(0)}_{\mu\nu}. \tag{3.11}$$

## 3.2 Lie algebraic deformations and the example of $\kappa$-deformation

Next class of space-time deformations is described by Lie-algebraic relation

$$[\widehat{x}_\mu, \widehat{x}_\nu] = \frac{1}{\kappa} \theta^{(1)\rho}_{\mu\nu} \widehat{x}_\rho, \tag{3.12}$$

where the constant dimensionless coefficients $\theta^{(1)\rho}_{\mu\nu}$ should satisfy the Jacobi identities

$$\theta^{(1)\rho}_{\mu\nu} \theta^{(1)}{}_{\rho\lambda}{}^\tau + \text{cycl}(\mu, \nu, \lambda) = 0. \tag{3.13}$$

The introduction of noncommutative translations

$$\widehat{x}_\mu{}' = \widehat{x}_\mu + \widehat{a}_\mu, \tag{3.14}$$

which preserve unchanged the relations (3.12) leads to the following conditions

$$[\widehat{a}_\mu, \widehat{a}_\nu] = \frac{1}{\kappa} \theta^{(1)\rho}_{\mu\nu} \widehat{a}_\rho \qquad [\widehat{a}_\mu, \widehat{x}_\nu] = 0. \tag{3.15}$$

The addition relation (3.14) can be represented as a primitive coproduct of the noncommutative translation generators $\widehat{x}_\mu$

$$\Delta(\widehat{x}_\mu) = \widehat{x}_\mu \otimes 1 + 1 \otimes \widehat{x}_\mu, \tag{3.16}$$

where we identify in (3.16) $\widehat{x}_\mu \otimes 1 \equiv \widehat{a}_\mu$ and $1 \otimes \widehat{x}_\mu = \widehat{x}_\mu$.

The relation (3.16) for particular choices of the structure constant $\theta^{(1)\rho}_{\mu\nu}$ can be extended to the quantum Poincaré group $\mathscr{P}^q_4 = (\widehat{x}_\mu, \widehat{\Lambda}_\mu{}^\nu)$, where $\widehat{\Lambda}_\mu{}^\nu$ describes the deformed Lorentz group parameters. We add that almost complete classification of the Hopf-algebraic deformations of Poincaré group has been presented by Podleś and Woronowicz ([29]; see also [30]). However, only part of the Lie-algebraic deformations of Poincaré group can be described by the introduction of the twist factor $F$ [31]. In particular first and the most familiar Lie-algebraic deformation of Poincaré symmetries, the $\kappa$-deformation, can not be described by a twist[5]. The $\kappa$-deformed Minkowski space [35]–[37]

$$[\widehat{x}_j, \widehat{x}_i] = \frac{i}{\kappa} \widehat{x}_0, \qquad [\widehat{x}_i, \widehat{x}_j] = 0, \tag{3.17}$$

is enlarged to $\kappa$-deformed Poincaré group by supplementing of the following deformed Lorentz sector [38]

$$[\Lambda^\mu{}_\nu, \Lambda^\rho{}_\tau] = 0 \qquad \Lambda^\mu{}_\nu \Lambda^\nu{}_\tau = \delta^\mu_\tau \tag{3.18}$$

$$[\Lambda^\mu{}_\nu, x_\rho] = -\frac{i}{\kappa} \left[ (\Lambda^\mu{}_0 - \delta^\mu_0)\Lambda_{\rho\nu} + (\Lambda^0{}_\nu - \delta^0_\nu)\delta^\mu{}_\rho \right]. \tag{3.19}$$

For the physical applications more useful appears to be the dual picture of the $\kappa$-deformed quantum Poincaré algebra [9, 36]. We shall present it below in so-called bicrossproduct form[6]

a) algebraic sector ($P_\mu = (P_0, P_i)$, $M_{\mu\nu} = (M_i, N_i)$)

$$[M_{\mu\nu}, M_{\rho\tau}] = i(\eta_{\mu\tau} M_{\nu\rho} - \eta_{\mu\rho} M_{\nu\tau} + \eta_{\nu\rho} M_{\nu\tau} - \eta_{\nu\tau} M_{\mu\rho})$$

$$[M_i, P_j] = i\varepsilon_{ijk} P_k \qquad [M_i, P_0] = i P_i$$

$$[N_i, P_j] = i\delta_{ij} \left[ \frac{\kappa}{2} \left(1 - e^{-\frac{2P_0}{\kappa}}\right) + \frac{1}{2\kappa}\vec{P}^2 \right] - \frac{i}{\kappa} P_i P_j$$

$$[N_i, P_0] = i P_i. \tag{3.20}$$

b) coalgebraic sector

$$\Delta(P_0) = P_0 \otimes 1 + 1 \otimes P_0$$

$$\Delta(P_i) = P_i \otimes e^{-\frac{P_0}{\kappa}} + 1 \otimes P_i$$

$$\Delta(M_i) = M_i \otimes 1 + 1 \otimes M_i$$

---

[5] We add here that $\kappa$-deformation has been described by various twists (see e.g. [32]–[34]) only in the approximate sense.

[6] For the definition of bicrossproduct Hopf algebra see [13]. In bicrossproduct basis it is technically easier to construct the dual pair of quantum Poincaré group and quantum Poincaré algebra.

$$\Delta(N_i) = N_i \otimes e^{-\frac{P_0}{\kappa}} + 1 \otimes N_i - \frac{1}{\kappa} \varepsilon_{ijk} M_j \otimes P_k \qquad (3.21)$$

c) coinverse (antipode)

$$S(P_0) = -P_0 \qquad S(M_i) = -M_i$$
$$S(P_i) = -e^{\frac{P_0}{\kappa}} P_i \qquad S(N_i) = e^{-\frac{P_0}{\kappa}} N_i + \frac{1}{\kappa} \varepsilon_{ijk} e^{\frac{P_0}{\kappa}} P_j M_k \,. \qquad (3.22)$$

Two Casimirs of Poincaré algebra, describing rest mass and spin, are described by

a) $\kappa$-deformed mass Casimir [36, 37]

$$C_2^{(\kappa)} = 2\kappa \left(\sinh \frac{P_0}{2\kappa}\right)^2 - \vec{p}^{\,2} e^{-\frac{P_0}{\kappa}} \qquad (3.23)$$

b) $\kappa$-deformed relativistic spin Casimir [39, 40]

$$C_4^{(\kappa)} = \left(\cosh \frac{P_0}{\kappa} - \frac{\vec{P}^2}{4\kappa^2}\right) W_0^2 - \vec{W}^{\,2} \,, \qquad (3.24\text{a})$$

$$W_0 = \vec{P} \cdot \vec{M} \qquad W_i = \kappa\, M_i \sinh \frac{P_0}{\kappa} + \varepsilon_{ijk}\, P_j M_k \,. \qquad (3.24\text{b})$$

We see from (3.20–3.21) that only the subalgebra $O(3) \oplus R$, describing space rotations and time translations, remains classical as a Hopf algebra. The deformation of the coproduct for the three-momentum operator (see (3.21)) leads to the modification of the Abelian addition law of the three-momenta.

## 3.3 Quadratic deformations of Minkowski space

Such deformations are described by the algebra

$$[\widehat{x}_\mu, \widehat{x}_\nu] = \theta^{(2)\rho\tau}_{\mu\nu} \widehat{x}_\rho \widehat{x}_\tau \,, \qquad (3.25)$$

which does not contain any dimensionfull parameter (see e.g. [14, 15]). A particular example of such deformation is obtained if we deform the standard commutator by $q$-deformed one (see e.g. [10])

$$[A, B] = AB - BA \to [A, B]_q = AB - qBA \,. \qquad (3.26)$$

The noncommutative translations (3.14) which preserve the commutation relations (3.25) should satisfy the following relations

$$[\widehat{a}_\mu, \widehat{a}_\nu] = \theta^{(2)\rho\tau}_{\mu\nu} \widehat{a}_\rho \widehat{a}_\tau \,, \qquad (3.27\text{a})$$
$$[\widehat{a}_\mu, \widehat{x}_\nu] = \theta^{(2)\rho\tau}_{\mu\nu} \widehat{a}_\rho \widehat{x}_\tau \,. \qquad (3.27\text{b})$$

From (3.27b) follows that the shift (3.14) can not be described by the coproduct formula (3.16) unless we introduce the braided tensor product with its symmetry properties defined as follows

$$\widehat{x}_\mu \widetilde{\otimes} \widehat{x}_\nu = \theta^{(2)\rho\tau}_{\mu\nu} \widehat{x}_\rho \widetilde{\otimes} \widehat{x}_\tau \,, \tag{3.28}$$

where $\theta^{(2)\rho\tau}_{\mu\nu}$ plays the role of braid factor.

In such a case the noncommutative translations (3.27a–3.27b) can be described again by the classical addition formula

$$\Delta(\widehat{x}_\mu) = \widehat{x}_\mu \widetilde{\otimes} 1 + 1 \widetilde{\otimes} \widehat{x}_\mu \,. \tag{3.29}$$

Due to the appearance of braided tensor product the quadratic deformation of space-time is described by the translation sector of braided quantum group [13, 41]. In such a case the deformation does change besides the algebra and coalgebra relations as well the definition of tensor product. We add that the braided quantum Poincaré symmetries leading to the choice of quadratic deformation of space-time coordinates (e.g. the example of so-called $q$-deformation of Poincaré algebra) has been investigated already in early nineties [10, 42, 43].

## 4 Deformation of classical and quantum mechanics

The models of classical and quantum classical mechanics formulated in phase space $Y_A = (x_\mu, p_\mu)$ are specified if we provide

a) Basic symplectic structure described by Poisson brackets (PB)

$$\omega_{AB}(Y) = \{Y_A, Y_B\} \,. \tag{4.1}$$

Antisymmetric two-tensor (4.1) defines the two-form $\omega_2 = \omega^{AB} \, dY_A \wedge dY_B$ which due to the Jacobi identities is closed

$$d\omega_2 = 0 \,. \tag{4.2}$$

In quantized dynamical models the symplectic structure determines the structure of equal time (ET) commutators in quantum phase space $\widehat{Y}_A = (\widehat{x}_\mu, \widehat{p}_\mu)$

b) The Hamiltonian $H(Y_A)$ which in Hamiltonian framework specifies the dynamics. In order to introduce the Lagrangean one should introduce the Liouville one-form $\Omega_1 = \Omega^A \, dY_A$ where

$$\omega_2 = d\Omega_1 \,, \tag{4.3}$$

Lagrangean density in phase space is defined as follows

$$\mathscr{L} = \Omega^A \, \dot{Y}_A - H(Y_A) \,. \tag{4.4}$$

The deformed relativistic phase space should incorporate in its prequantized version the PB of space-time coordinates corresponding to the noncommutativity relation (1.2). In both canonical and $\kappa$-deformed quantum deformations the fourmomentum generators are Abelian, i.e. we can assume in phase space that their P.B. counterpart looks as follows

$$\{p_\mu, p_\nu\} = 0. \tag{4.5}$$

The remaining "crossed" PB relations $\{x_\mu, p_\nu\}$ can be obtained in two ways:

(i) One can always add the "crossed" relations to the P.B. of coordinates (we shall consider here expicitly the P.B. counterparts of (1.1) and (3.17)) and to the relations (4.5) in algebraically consistent way (Jacobi identities). In such a way we need not to consider the possible quantum symmetries of the system, and we obtain a large class of possible PB structures [37, 44, 45].

(ii) One can derive uniquely the deformed phase space structure from the dual pair of Hopf algebras describing deformed symmetries: quantum Poincaré group and quantum Poincaré algebra. For particular deformations, if space-time translations and fourmomentum generators form respectively the pair of dual Hopf subalgebras $(H_x, H_p)$, the corresponding quantum phase space is described by so-called Heisenberg double construction [46, 47]. If $x \in H_x$ and $p \in H_p$, the quantum deformation of the Heisenberg algebra relation (we put $\hbar = 1$)

$$[\widehat{x}_\mu, \widehat{p}_\nu] = i\,\eta_{\mu\nu} \tag{4.6}$$

is given by the cross-multiplication formula (no summation on rhs!)

$$\widehat{x}_\mu \cdot \widehat{p}_\nu = \widehat{p}_\nu^{(1)} \langle \widehat{x}_\mu^{(1)}, \widehat{p}_\nu^{(2)} \rangle \widehat{x}_\mu^{(2)}, \tag{4.7}$$

where $\Delta(\widehat{x}_\mu) = \widehat{x}_\mu^{(1)} \otimes \widehat{x}_\mu^{(2)}$, $\Delta(\widehat{p}_\mu) = \widehat{p}_\mu^{(1)} \otimes \widehat{p}_\mu^{(2)}$, and $\langle x, p \rangle$ denotes the nondegenerate scalar product in $H_x \otimes H_p$. The standard dual bases in $H_x, H_p$ are described by the relations

$$\langle \widehat{x}_\mu, \widehat{p}_\nu \rangle = i\,\eta_{\mu\nu}. \tag{4.8}$$

Because $\widehat{x}_\mu$ and $\widehat{p}_\mu$ are the modules of deformed relativistic symmetries, the construction of phase space based on crossed relations (4.6) is covariant under the quantum symmetries.

If the Hopf subalgebras $H_x$ and/or $H_p$ can not be extracted respectively from quantum Poincaré group and/or quantum Poincaré algebra, the Heisenberg double construction should be applied to the whole dual pair of deformed Poincaré group and Poincaré algebras. Such generalized phase space will contain all noncommutative Poincaré group coordinates $(\widehat{x}_\mu, \widehat{\Lambda}_\mu{}^\nu)$ and generalized noncommutative momenta $(\widehat{p}_\mu, \widehat{m}_{\mu\nu})$. In such a case the phase space $(\widehat{x}_\mu, \widehat{p}_\mu)$ can not be considered separately from the angular momentum degrees of freedom $(\widehat{\Lambda}_\mu{}^\nu, \widehat{m}_{\mu\nu})$, and we get the deformation of dynamics which is formulated on 10-dimensional Poincaré group manifold (for undeformed case see e.g. [48]).

Now we shall consider our two specific deformations:

### 4.1 Canonical deformation

In canonical case we can not employ the construction of quantum phase space based on Heisenberg double formula, because in canonically deformed quantum Poincaré group the algebra of noncommutative translations is not closed. Due to the formula (3.5) and duality we obtain the following relation [49]

$$[\hat{x}_\mu, \hat{x}_\nu] = \frac{i}{\kappa^2} \left( \theta^{(0)}_{\mu\nu} - \theta^{(0)}_{\rho\tau} \widehat{\Lambda}^\rho_\mu \widehat{\Lambda}^\tau_\nu \right) \tag{4.9}$$

i.e. we can only define the $\theta$-deformed covariant phase space on the whole canonically deformed Poincaré group manifold.

In algebraic approach (see i)) we can however easily complete the phase space algebra by adding to the relations (1.1) and (4.4) the classical PB algebra relations

$$\{x_\mu, p_\nu\} = i\, \eta_{\mu\nu}\,. \tag{4.10}$$

The relations (1.1), (4.4), (4.10) are generated by the following symplectic form

$$\omega^{(0)}_{AB} = \frac{1}{\kappa^2} \theta^{(0)\nu}_\mu dp^\mu \wedge dp_\nu + dp^\mu \wedge dx_\mu\,. \tag{4.11}$$

It follows that (see (4.3))

$$\Omega^{(\theta)}_1 = \frac{1}{\kappa^2} \theta^{(0)\nu}_\mu p^\mu dp_\nu + p^\mu dx_\mu \tag{4.12}$$

and the counterpart of free particle Lagrangean in canonically deformed phase space looks as follows

$$\mathcal{L}^{(\theta)}_o = p^\mu \dot{x}_\mu + \theta^{\mu\nu} p_\mu \dot{p}_\nu - H(x_\mu, p_\mu)\,. \tag{4.13}$$

In D=2+1 the model leading to the first order action (4.13) has been firstly introduced in [50] (see also [51, 52]). The model (4.13) can be expressed however in terms of canonical variables $(X_\mu, p_\mu)$, where $X_\mu = x_\mu - \theta_\mu^{\ \nu} p_\nu$, because modulo total differential the Liouville one-form (4.12) can be written as describing the standard free classical particle

$$\Omega^{(\theta)}_1 = -X^\mu dp_\mu\,, \tag{4.14}$$

where $\{X_\mu, X_\nu\} = 0$ and $\{X_\mu, p_\nu\} = \eta_{\mu\nu}$. We add that more general framework for the classical mechanics in $\theta$-deformed space-time has been presented in [53].

### 4.2 $\kappa$-deformed classical and quantum mechanics

First studies of $\kappa$-deformed classical and quantum mechanics were presented in [37], where the algebraic approach (see i)) was used. In particular in [37] it was initiated the method of representing the $\kappa$-deformed phase spaces in terms of standard relativistic phase space variables. Such a way of describing $\kappa$-deformed phase geometry has been recently further developed (see e.g. [44, 45]) in the framework of so-called Doubly Special Relativity[7].

Here we shall describe the $\kappa$-covariant approach, obtained by application of formula (4.6) to the construction of the $\kappa$-deformed phase space. The $\kappa$-deformed dual Hopf subalgebras $H_x$ and $H_p$ take the following form

$$H_x: \quad [\widehat{x}_0, \widehat{x}_i] = \frac{i}{\kappa} \qquad [\widehat{x}_i, \widehat{x}_j] = 0$$

$$\Delta \widehat{x}_i = \widehat{x}_i \otimes 1 + 1 \otimes \widehat{x}_i \tag{4.15}$$

$$H_p: \quad [\widehat{p}_\mu, \widehat{p}_\nu] = 0$$

$$\Delta(\widehat{p}_0) = \widehat{p}_0 \otimes 1 + 1 \otimes \widehat{p}_0$$

$$\Delta(\widehat{p}_i) = \widehat{p}_i \otimes e^{-\frac{\widehat{p}_0}{\kappa}} + 1 \otimes \widehat{p}_i. \tag{4.16}$$

Using (4.7) one gets from (4.15–4.16) the following crossed relations [46, 47, 58]

$$[\widehat{x}_k, \widehat{p}_l] = i\,\delta_{kl} \qquad [\widehat{x}_0, \widehat{p}_l] = \tfrac{i}{\kappa}\widehat{p}_l$$

$$[\widehat{x}_k, \widehat{p}_0] = 0 \qquad [\widehat{x}_0, \widehat{p}_0] = -i. \tag{4.17}$$

The phase space relations (4.17) are generated by the following symplectic form:

$$\omega_2^{(\kappa)} = dp^\mu \wedge dx_\mu - \frac{1}{\kappa}(x_i dp_i \wedge dp_0 + p_i dx_i \wedge dp_0). \tag{4.18}$$

One gets

$$\Omega_1^{(\kappa)} = p^\mu dx_\mu - \frac{1}{\kappa}(\vec{p}\vec{x})dp_0. \tag{4.19}$$

The $\kappa$-deformed Lagrangean based on one-form (4.18) has been recently considered in [59]. It should be added that if we introduce the variables

$$X_0 = x_0 - \frac{1}{\kappa}\vec{p}\vec{x} \tag{4.20}$$

the one-form (4.19) can be transformed into the standard one, given by (4.14).

The property that deformed phase space can be embedded into the standard undeformed one tells us that the quantum deformation can be represented by the noncanonical transformations of the standard phase space which makes the phase space curved [60, 61]. In particular it has been shown [62] that the classical curved structure in fourmomentum space due to $\kappa$-deformation can be described by de-Sitter geometry. We conclude that the formulation of deformed dynamics in curved phase space is provided by the well-known Hamiltonian formalism with noncanonical symplectic structure

$$Y_A = \omega_{AB} \frac{\partial H}{\partial Y_B}. \tag{4.21}$$

---

[7]Doubly Special Relativity (DSR) has been initiated by G. Amelino-Camelia [54]; see also [55, 56]. In DSR it is considered the extension of Special Relativity framework introducing besides the light velocity a second fundamental parameter describing mass or length. The $\kappa$-deformation of relativistic symmetries and the theory of nonlinear realizations of classical Poincaré symmetries are two mathematically the most complete realizations of DSR ideas [57].

# 5 Deformed noncommutative field theory

## 5.1 General considerations

In the description of perturbative quantum field theory the main building blocks are quantum free fields. In this section we shall consider the quantum deformations of free fields.

The standard (undeformed) free scalar field is described by the formula

$$\begin{aligned}
\phi_0(x) &= \frac{1}{(2\pi)^4} \int d^4p\, A(p)\delta(p^2 - m^2)e^{ipx} \\
&= \frac{1}{(2\pi)^3} \int \frac{d^3\vec{p}}{2\omega(\vec{p})} \left(a^+(\vec{p})\, e^{i(\vec{p}\vec{x}-\omega t)} + a(\vec{p})e^{-(\vec{p}\vec{x}-\omega t)}\right), \quad (5.1)
\end{aligned}$$

where $\omega(\vec{p}) = (p^2 + m^2)^{\frac{1}{2}}$ and $a^+(\vec{p}) = A(\omega(\vec{p}), \vec{p})$, $a(\vec{p}) = A(-\omega(\vec{p}), -\vec{p})$. For undeformed classical fields the exponentials $e^{ipx}$ and $A(p)$ are usual functions; for quantum free fields the functions $A(p)$ (i.e. $a(\vec{p})$ and $a^+(\vec{p})$) are becoming field oscillators

$$[a(\vec{p}), a^+(\vec{q})] = 2\omega\, \delta^3(\vec{p} - \vec{p}') \tag{5.2a}$$
$$[a(\vec{p}), a(\vec{q})] = [a^+(\vec{p}), a^+(\vec{q})] = 0. \tag{5.2b}$$

The noncommutative modification of the free fields (5.1) can be obtained by the superposition of the following three operations:

i) We replace the classical Minkowski space-time coordinates by the noncommutative ones (see (1.2)). The algebra of functions on noncommutative space-time can be represented homomorphically as the algebra of functions on classical Minkowski space with the multiplication described by the $\star$-product. As we mentioned already such realization is also denoted as the Weyl map.

ii) We modify the algebraic structure of the algebra of Fourier field components $A(p)$. Even in the case of classical deformed field theory the standard functions $A(p)$ are becoming the noncommuting operators. There are two ways of introducing such a modification:

1) One modifies the notion of the multiplication of the operators $A(p) = a(\vec{p})$, $a^+(\vec{p})$, e.g.

$$a(\vec{p}) \cdot a(\vec{q}) \longrightarrow a(\vec{p}) \circ a(\vec{q}), \tag{5.3}$$

where ∘-multiplication usually can be expressed in momentum space as the standard multiplication with nonlocal kernel

$$a(\vec{p}) \circ a(\vec{q}) = \int d^3\vec{r}\, d^3\vec{s}\, K(\vec{p}, \vec{q}; \vec{r}, \vec{s}) a(\vec{r}) a(\vec{s}) \,. \tag{5.4}$$

One can further postulate that for quantized fields the deformation of the algebra (5.2a) is described only by the replacement (5.3) of the multiplication rule, i.e. the deformed field oscillator algebra is described by the commutation relations analogous to (5.2a–5.2b) ($[A, B]_\circ = A \circ B - B \circ A$)

$$[\widehat{a}(\vec{p}), \widehat{a}^+(\vec{q})]_\circ = 2\,\Omega(\vec{p})\,\delta^3(\vec{p} - \vec{q})\,, \tag{5.5a}$$
$$[a(\vec{p}), a(\vec{q})]_\circ = [a^+(\vec{p}), a^+(\vec{q})]_\circ = 0\,. \tag{5.5b}$$

The function $\Omega(\vec{p})$ may be different from $\omega(\vec{p})$ if the mass-shell condition $p^2 - m^2 = 0$ is modified by deformation.

2) The modification of relations (5.2a) can be obtained as well by the introduction of braided form of the commutator

$$[A(p), A(q)] \longrightarrow [\widetilde{A}(p), \widetilde{A}(q)]_{br} :$$
$$:= \widetilde{A}(p)\,\widetilde{A}(q) - \int d^4s\, d^4t\, R_\xi(p, q, s, t) \widetilde{A}(s)\,\widetilde{A}(t)\,, \tag{5.6}$$

where the braid factor $R_\xi$ satisfies suitable additional conditions (see e.g. [63]), in particular in "no deformation" limit $\xi \to 0$ becomes the local flip operator.

$$R_\xi(p, q; s, t) \xrightarrow[\xi \to 0]{} \delta^4(p - t)\delta^4(q - s)\,. \tag{5.7}$$

By deformation procedure the classical free fields are becoming the braided free fields, with its Fourier modes $A(p)$ braided-commutative

$$[\widetilde{A}(p), \widetilde{A}(q)]_{br} = 0\,. \tag{5.8}$$

iii) Third possibility of deforming the algebra of free fields is generated in the product of two deformed fields by possible lack of commutativity between the noncommutative exponentials and field oscillators[8]

$$e^{ip\widehat{x}}\,\widehat{A}(q) \neq \widehat{A}(q)\, e^{ip\widehat{x}}\,. \tag{5.9}$$

Such noncommutativity is justified if we observe that the algebra of noncommutative exponentials and field oscillators describe two different representation modules of deformed Poincaré algebra. It is known from the theory of tensor representations of quasitriangular quantum groups that between two factors in tensor product the quantum deformation introduces nontrivial braiding, determined by the form of universal $R$-matrix (see (2.16)).

---

[8]Braiding iii) has been e.g. described in [20] (see e.g. last formula in (4.6))

It should be added that the quantum modification of formula (5.1) should additionally take into consideration the modification of mass-shell condition, changing the classical relativistic dispersion relation $p_0 = \omega(\vec{p})$ and the numerical factor in the relation (5.2a) (see (5.5a)). In general case the integration measure $d^4k$ should be also modified, in a way preserving the invariance under the deformed Lorentz transformations in four-momentum space.

## 5.2 Canonically deformed quantum free fields

For canonical deformation the noncommutativity of coordinates (1.1) is not accompanied by the modification of mass-shell condition, and because the Lorentz transformations in algebraic sector are not modified, the integration measure $d^4k$ in fourmomentum space remains unchanged.

We recall here that the canonical deformation of relativistic symmetries (Poincaré group and Poincaré algebra) are described by a twist factor (3.6), what permits in principle to determine all three factors i), ii), iii) entering into the deformation of free fields. The deformation i) due the noncommutativity (1.1) of local field arguments is unique, and obtained by the replacement of standard multiplication by the nonlocal Moyal-Weyl star product of fields (3.9). However concerning the choices of deformations ii) and iii) it has not been achieved an agreement in the literature (see e.g. [20]). I shall describe the choice which may be considered priviledged ([65]; see also [64]).

Let us consider firstly the factor i), the noncommutative deformation of the classical space-time coordinates. From (1.1)) follows the formula

$$e^{ip\widehat{x}} \cdot e^{iq\widehat{x}} = e^{i(p+q)\widehat{x}} e^{-\frac{i}{2}p\,\theta\,q} \neq e^{iq\widehat{x}} \cdot e^{ip\widehat{x}}. \tag{5.10}$$

If we assume that an additional pair of coordinates $\widehat{y}_\mu$ satisfies the algebra

$$[\widehat{x}_\mu, \widehat{y}_\nu] = \frac{i}{\kappa^2}\theta^{(0)}_{\mu\nu}, \tag{5.11}$$

$$[\widehat{y}_\mu, \widehat{y}_\nu] = \frac{i}{\kappa^2}\theta^{(0)}_{\mu\nu}, \tag{5.12}$$

with the relation (5.11) introducing suitable braiding factor between two Poincaré algebra modules $\widehat{X} = (\widehat{x}_\mu)$ and $\widehat{Y} = (\widehat{y}_\mu)$, we obtain the following generalization of (5.10)

$$e^{ip\widehat{x}} \cdot e^{iq\widehat{y}} = e^{i(p\widehat{x}+q\widehat{y})} e^{-\frac{i}{2}p\,\theta\,q}. \tag{5.13}$$

The multiplication rule (5.13) in the limit $\widehat{y}_\mu \to \widehat{x}_\mu$ leads consistently to (5.10).

The set of algebraic relation (5.13) is represented on the space of classical exponentials if we introduce the Weyl map

$$e^{ip\widehat{x}} \cdot e^{iq\widehat{y}} \xrightarrow{W} e^{ipx} \star_\theta e^{iqy}, \tag{5.14}$$

with $\star_\theta$ - star product defined by the formula (3.9) extended to bilocal products:

$$f(x) \star_\theta g(y) = f(x) \exp\left\{\frac{i}{\kappa^2} \overleftarrow{\partial}_x^\mu \theta_{\mu\nu}^{(0)} \overrightarrow{\partial}_y^\nu\right\} g(y). \tag{5.15}$$

The relation (5.15) for general twist factor $\widehat{F}$ corresponds to an extension of the formula (3.7) to the bilocal $\star_\theta$-product of functions $f(x), g(y)$.

In order to obtain second factor (see ii)) defining the algebra of deformed free quantum fields we can apply the relation (3.7) as well to the description of the deformed multiplication (see (5.3)) of functions describing field quanta in momentum space. We get

$$\begin{aligned}\widehat{A}(p) \circ \widehat{A}(q) \xrightarrow{W} A(p) \star_\theta A(q) &= A(p) \exp\left\{\frac{i}{\kappa^2} p^\mu \theta_{\mu\nu}^{(0)} q^\nu\right\} A(q) \\ &= \exp\frac{i}{\kappa^2} p^\mu \theta_{\mu\nu}^{(1)} q^\nu A(p) A(q),\end{aligned} \tag{5.16}$$

where $p_0 = \omega(\vec{p})$, $q_0 = \omega(\vec{q})$. The formula (5.16) describes in unified way the deformed products of the creation and annihilation operators $a(\vec{p}), a^+(\vec{p})$. Further, in accordance with requirement that the whole deformation is the modification of the multiplication rule, we postulate the standard oscillator algebra relations (5.5a–5.5b), with $\Omega(\vec{p}) = \omega(\vec{p})$.

The remaining part of the deformation is the explicit description of the noncommutativity (5.9) via braid factor (see iii)). For that purpose we should use formula (2.16) where $\widehat{R}$ is given by formula (2.18), with $F$ given by (3.6). Because $F_\theta^T = F_\theta^{-1}$ we obtain that

$$R_\theta = F_\theta^{-2} = \exp\left\{-\frac{2i}{\kappa^2} \theta_{\mu\nu}^{(0)} P^\mu \otimes P^\nu\right\} \tag{5.17}$$

and the twist factor causing inequality (5.9) leads to the equation

$$e^{ipx} A(p) = e^{-\frac{2i}{\kappa^2} p^\mu \theta_{\mu\nu}^{(0)} q^\nu} A(p) e^{ipx}. \tag{5.18}$$

Let us take now into consideration all three deformation factors described by the formulae (5.15), (5.16) and (5.18) and substitute into the product of fields

$$\phi_0(\widehat{x})\phi_0(\widehat{y}) \xrightarrow{W} \phi_0(x) \circledast_\theta \phi_0(y). \tag{5.19}$$

One can check that the three phase factors in momentum space following from (5.15), (5.16) and (5.18) do cancel. We obtain therefore in effect the multiplication rule of standard free fields, i.e.

$$\phi_0(x) \circledast_\theta \phi_0(y) = \phi_0(x) \phi_0(y), \tag{5.20}$$

because the total multiplication $\circledast_\theta$ of canonically deformed fields which is composed by three operations given by (5.15), (5.16) and (5.18), gives identity. Calculating in quantum case the commutator, we obtain

$$[\phi_0(x), \phi_0(y)]_\circledast = [\phi_0(x), \phi_0(y)] =$$

$$= \tfrac{1}{i}\Delta(x-y; m^2) = \tfrac{1}{(2\pi)^3}\int d^4p\, \varepsilon(p_0)\delta(p^2-m^2)e^{ipx}\,. \tag{5.21}$$

We see therefore that we obtained a sequence of deformations, which cancel each other and lead to standard free quantized fields - with nonlocal multiplication of space-time functions, modified field oscillators algebra (i.e. introducing modified statistics [66]) and "compensation" of both mentioned effects by suitably chosen braid factor.

### 5.3 $\kappa$-deformed quantum free fields

The $\kappa$-deformation of free quantum fields is more complex in comparison with canonical deformation described in last subsection. In particular

1) The mass-shell condition $p^2 - m^2 = 0$ is modified (see (3.23)) and it depends on the chosen $\kappa$-deformed Poincaré basis. This modification changes the relativistic energy-momentum relation $p_\circ = \omega(\vec{p})$ as well as the functional factor in the algebra of field oscillators (see (5.2a)).

2) The Lorentz-invariant measure is changed; in bicrossproduct basis it is given by the following modification also [67]

$$d^4k \to [d^4k]_\kappa = e^{\frac{3p_\circ}{\kappa}}\, d^4k\,. \tag{5.22}$$

3) Unfortunately we do not know the universal $\widehat{R}$-matrix for the $\kappa$-deformed Poincaré algebra. Because the classical $r$-matrix describing $\kappa$-deformations satisfies modified Yang-Baxter equation, we know for sure that the $\kappa$-deformation of Poincaré algebra can not be described by the cocycle twist, satisfying the relations (2.22). If we postulate however the existence of similarity map, transposing the coproducts (see (2.19)) it was argued recently that such a cochain twist $\widehat{F}$ exists only in the category of quasi-Hopf-alebras [68, 69].

Recently there were described $\kappa$-deformed free quantum fields with modified algebra of oscillators [24, 70, 71] but they correspond only to the approximate description of $\kappa$-deformation. Indeed, further it has been shown [72] that the deformation of oscillators algebra proposed firstly in [70] can be obtained by twist deformation of the classical Poincaré algebra by the following twist factor

$$F_\kappa = \exp\frac{i}{\kappa}D \wedge P_0\,, \tag{5.23}$$

where $D$ is the dilatation operator which defines the Weyl extension of Poincaré algebra $(M_{\mu\nu}, P_\mu) \to (M_{\mu\nu}, P_\mu, D)$. Because the carrier algebra of the twist (5.23) does not belong to Poincaré algebra, it leads to unwanted difficulties, such as nonclosure of the Poincaré coalgebra (the coproducts of boost generators depend on $D$).

According to our opinion the formulation of the $\kappa$-deformed field theory which is covariant under arbitrary change of the $\kappa$-deformed frames described by Hopf algebra (3.20–3.22) is still the challenge for the future.

# 6 Quantum deformations of Einstein gravity

Quantum gravity, which should unify the geometric framework of General Relativity with the principles of quantum theories is now the most fascinating and unsolved subject of fundamental physics. Direct quantization of Einstein-Hilbert action (Einstein gravity) using perturbative methods leads to nonrenormalizable divergencies and conceptual problems. If the perturbative methods should better succeed, it is necessary to introduce new perturbative expansions, describing for example at lowest order the Schwarzschild solution[9]. One seeks usually the removal of nonrenormalizability by embedding gravity in larger dynamical systems, like supergravity or superstring theory. Other way which has been also studied recently is to consider Einstein action on modified, noncommutative space-time as nonlinear noncommutative field theory.

If the noncommutative structure of space-time can be derived by the introduction of twist factor $F$ defining twisted Poincaré-Hopf algebra, one can pass from the standard (pseudo)Riemannian geometry of general relativity to deformed gravity by introduction of suitable $\star$-products (see e.g. [74]–[76]). Out of possible three sources of deformations in QFT, considered in Sect. 5 (see i), ii) and iii)) only the first one (see i)), due to the replacement of classical spacetime by noncommutative set of coordinates, has been taken into consideraton. In such a framework the algebra of field oscillators describing gravitons is still described by standard undeformed oscillator algebra.

The deformed Einstein-Hilbert action is given by the nonlocal expression

$$\widetilde{S}^{EH} = \int d^4x\, \varepsilon^{\mu\nu\rho\tau}\, \varepsilon_{abcd}\, e^a_\mu \star \widetilde{R}^{bc}_{\nu\rho} \star e^d_\tau, \tag{6.1}$$

where

$$\widetilde{R}^{bc}_{\nu\rho} = \partial_{[\mu}\widetilde{\omega}^{\,bc}_{\ \nu]} + \widetilde{\omega}^{\,bf}_{\ [\mu} \star \widetilde{\omega}^{\ c}_{f\ \nu]} \tag{6.2}$$

and spin connection $\widetilde{\omega}^{bc}_\nu$ can be expressed as nonlocal function of vierbeins obtained by solving the equation $\widetilde{T}^a_{\mu\nu} = 0$ which describes vanishing noncommutative torsion

$$\widetilde{T}^a_{\mu\nu} = \partial_{[\mu}e^a_{\nu]} - \omega^a_{\ b[\mu} \star e^b_{\ \nu]}. \tag{6.3}$$

If we introduce the $\star$-deformed determinant $\det_\star e$ of the vierbein fields matrix, the action for deformed Einstein gravity can be written as well as the $\star$-deformation of standard form of the Einstein-Hilbert action

$$\widetilde{S}^{EH} = \int d^4x\, \varepsilon^{\mu\nu\rho\tau}\, \det{}_\star e \star e^a_\mu \star R_{ab\rho\tau} \star e^b_\tau. \tag{6.4}$$

The introduction of $\star$-multiplication permits to introduce $\star$-deformed differential calculus. We characterize the noncommutative deformation of general curved

---

[9] In standard perturbation theory the Schwarzschild solutions are described by infinite summ of tree graphs [73].

manifold by $\star$-deformed Cartan equations describing the deformed torsion and deformed curvature (see the formulae (6.3), (6.2)). The only local quantities are vierbeins $e^a_{\ \mu}$; the composite objects like spin connection $\widetilde{\omega}^{ab}_{\ \ \mu}$ or metric tensor $\widetilde{g}_{\mu\nu}$ given by formula

$$\widetilde{g}_{\mu\nu} = \frac{1}{2}\left(e^a_{\ \mu} \star e_{a\nu} + e^a_{\ \mu} \star e_{b\mu}\right) \tag{6.5}$$

are nonlocal. The deformed diffeomorphisms $\delta^\star_\xi = \xi^\mu_\star \partial_\mu$ satisfy the braided commutation relations (see (2.17)); $\widehat{R} = R^{(1)} \otimes R^{(2)}$

$$\delta^\star_\xi \delta^\star_\eta - \delta^\star_{(R^{(1)}\eta)}\delta^\star_{(R^{(2)}\xi)} \equiv \left[\delta^\star_\xi, \delta^\star_\eta\right]_\star = \delta_{(\xi\times\eta)_\mu \partial^\mu}, \tag{6.6}$$

where $(\xi\times\eta)_\mu = \xi^\nu \star \partial_\nu\eta_\mu - \eta^\nu \star \partial_\nu\xi_\mu$. The composition of two diffeomorphism is modified and described by twisted coproduct $\Delta^F(\delta^\star_\xi)$ what leads to nonAbelian modification of Leibniz rule

standard Leibniz          twisted Leibniz
rule                     rule

$$\Delta^{(0)}(\delta_\xi) = \delta_\xi \otimes 1 + 1 \otimes \delta_\xi \longrightarrow \Delta^F(\delta^\star_\xi) = \delta^{\star(1)}_\xi \otimes \delta^{\star(2)}_\xi. \tag{6.7}$$

The deformed Riemannian geometry has been recently studied for arbitrary cocycle twist describing triangular Poincaré-Hopf algebra [76], however the explicit calculations of quantum corrections following from actions (6.1) or (6.4) has been studied mostly for the canonical twist (3.6). If we expand the deformed Riemannian scalar curvature $R^\star = e_a^{\ \mu} \star R^{ab}_{\mu\nu} \star e_b^{\ \nu}$ into the power serie in canonical deformation parameters $\theta^{(0)}_{\alpha\beta}$

$$R^\star = R_{(0)} + \frac{1}{\kappa^2}\theta^{(0)}_{\alpha\beta}R^{\alpha\beta} + \frac{1}{\kappa^4}\theta^{(0)}_{\alpha\beta}\theta^{(0)}_{\gamma\delta}R^{\alpha\beta\gamma\delta} + \mathcal{O}(\frac{1}{\kappa^6}) \tag{6.8}$$

it has been shown [77, 78] that the first correction, proportional to $\frac{1}{\kappa^2}$ vanishes. Using the technique of Seiberg-Witten deformation map analogous conclusion about the first order correction has been obtained for Lie-algebraic deformations [79], in particular for the $\kappa$-deformation. It can be added that for the case of arbitrary coordinate-dependent deformation parameter $\theta_{\alpha\beta}(\widehat{x})$ it has been recently argued [80] that the first order correction is not be vanishing.

## 7 Final remarks

The noncommuttive space-time (1.2) and consistent replacement of standard Poincaré symmetries by quantum ones are the starting points of our approach. The modification introduced by noncommutativity should be present at ultra-short Planckian distances, which can be explored only by energies of probing particles corresponding to Planck mass $M_p \simeq 10^{19} GeV$. We recall that the scale of energies which is achieved in LHC is of order $10\,TeV \simeq 10^4 GeV$ i.e. we see that the particle accelerators are not able to check directly the quantum gravity effects. The only chance is to go beyond Earth with experimental set-ups and study the astrophysical effects, in particular the consequences of the processes at very early "quantum" stage of the Universe.

The mathematical framework presented in this lecture has various alternatives. Besides the loop quantum gravity (LQG) approach there are considered discrete versions of gravity models (see e.g. [81, 82]) and the spectral point of view on noncommutativity in gravity presented by A. Connes and his collaborators (see e.g. [83, 84]). In last approach the total space in elementary particle physics is the product of 4-dimensional Riemannian space-time and finite nonAbelian algebra $F$ (see [10]), reflecting the internal symmetry structure of Standard Model. The algebra $F$ is a finite noncommutative geometry, which can be represented by algebra of matrices. In order to obtain the quantum gravity effects at very high energies the authors conjecture in somewhat poetic way that also the gravitational sector becomes noncommutative [[84], p.18]. *"The small amount of noncommutativity encoded in finite geometry $F$ ... will gradually creep in and invade the whole algebra of coordinates which will become a huge matrix algebra at Planck scale."*

An important point in studying the noncommutative structures is to determine the measure of noncommutativity which should be a dynamical quantity. One can conjecture that the description of noncommutative gravity is incomplete without some dynamical equations determining the function $\theta_{\mu\nu}(\widehat{x})$ in (1.2)[11]. In dynamical Seiberg-Witten model of noncommutative $D$-brane coordinates described by the end points of $D=10$ strings the noncommutativity parameter is a function of dynamical geometric tensor fields $B_{\mu\nu}$, which enter into metric description of $D=10$ gravity. By analogy, even in pure quantum gravity described via vierbeins by deformed Einstein equations, there should be another equation, determining dynamically the local noncommutativity measure $\theta_{\mu\nu}(x)$

$$\theta_{\mu\nu}(x) = \theta_{\mu\nu}[e^a_\mu; x]. \tag{7.1}$$

If noncommutative approach to quantum gravity aspires to a closed effective framework, without such additional equations permitting to determine dynamically the "amount of noncommutativity" the formulation is not complete.

## Acknowledgments

Valuable discussions with Mariusz Woronowicz are acknowledged. The author would like to thank the organizers of 50-th Cracow Summer School of Physics, in particular Michał Praszałowicz, for invitation and warm hospitality in Zakopane. The paper has been supported by Ministry of Science and Higher Education grant NN 202-318534.

---

[10]$F$ is not related with twist factor; we simply use original Connes notation.
[11]Such a point of view was also advocated by S. Doplicher.